\documentclass[twocolumn]{article}
\usepackage{icqproc,epsf}

\newcommand\Q{\vec{\rm Q}}
\newcommand\R{\vec{\rm R}}

\def\Eunrel{{E^{\rm un}_{\rm rel}}}
\def\Erel{{E_{\rm rel}}}

\def\OMIT#1{{}}

\begin{document}
\thispagestyle{empty}

\twocolumn[
\vspace*{30mm}
\begin{LARGE} 
\begin{center}
Combined energy -- diffraction data refinement of decagonal AlNiCo
\end{center}
\end{LARGE}

\begin{large}
\begin{center} 
M. Mihalkovi\v{c}$^{1,3}$, C. L. Henley$^{2}$ and M. Widom$^{1}$
\end{center}
\end{large}

\begin{footnotesize}
\begin{it}
\begin{center}
$^{1}$Physics Dept., Wean Hall, Carnegie-Mellon University, Pittsburgh PA 15213 USA\\
$^{2}$Physics Dept., Clark Hall, Cornell University, Ithaca NY 14853 USA\\
$^{3}$Institute of Physics, Slovak Academy of Sciences, Bratislava, Slovakia\\
\end{center}
\end{it}
\end{footnotesize}
\begin{footnotesize}
\begin{center}
%
%
\date{\today}
%
%
\end{center}
\end{footnotesize}

\begin{small}
\hrule\vspace{3ex}
\begin{minipage}{\textwidth}
{\bf Abstract}\vspace{2ex}\\
\hp 
We incorporate realistic pair potential energies directly into a
non-linear least-square fit of diffraction data to quantitatively
compare structure models with experiment for the Ni-rich $d$(AlNiCo)
quasicrystal.  The initial structure models are derived from a few {\it a
priori} assumptions (gross features of the Patterson function) and the
pair potentials.  In place of the common hyperspace approach to the
structure refinement of quasicrystals, we use a real-space tile
decoration scheme, which does not rely on strict quasiperiodicity, and
makes it easy to enforce sensible local arrangements of the atoms.
Inclusion of the energies provides information complementary to the
diffraction data and protects the fit procedure from converging
on spurious solutions.  The method pinpoints sites which are likely to
break the symmetry of their local environment.
\vspace{2.5ex}

\end{minipage}\vspace{3ex}
\hrule
\end{small}\vspace{6ex}
]

\def\chem{\alpha}
\def\chemsite{\chi}
\def\orb{j}
\def\bxy{{b^{xy}}}
\def\bz{{b^{z}}}
\def\Esite{{E^{\rm site}}}
\def\Eorb{{E^{\rm orb}}}

\section{Introduction}
\label{sec:intro}
\hp

The decagonal AlNiCo system exhibits a variety of metastable and
stable phases, out of which the ``basic Ni-rich''
phase~\cite{Ni-basic} has attracted many detailed studies due to its
high structural quality.  Recently, two high-quality refinements have
been published based on single-crystal X-ray~\cite{CHS,TYT}.  At the
same time, a Monte Carlo method was developed~\cite{alnico} to {\it
predict} the same structure, approximating the Hamiltonian with pair
potentials, and using as inputs only the symmetry, the quasilattice
constant and the fact that the structure is strongly layered.  Here,
we present the first trial of a combined method which uses energy and
diffraction information to fix complementary features of the atomic
structure.

The refined structures of~\cite{CHS} and~\cite{TYT} agree on basic
features, but they differ in many important details.  The refinement
presented by CHS~\cite{CHS} exhibits 8\AA\ periodicity along the
vertical axis, lacks a 10-fold screw symmetry property, and does not
lend itself an easy interpretation in terms of a tiling geometry.  The
structure solution of \cite{TYT} has 4\AA\ periodicity, and can be
readily interpreted in terms of the Hexagon-Boat-Star (HBS) tiling
with edge length of $a_q\sim$6.5\AA.  The R-factors achieved are
comparable in the two cases.

Quasicrystals, as aperiodic structures, typically include
local patterns which are rare enough to be ill-determined
from diffraction, but common enough to have a significant
effect on physical properties (such as total energy
or conductivity). 
Thus the problem of structure prediction from energies becomes
entangled with structure fitting from diffraction.
We suspect that, in the future, the best 
quasicrystal structure fits 
will combine the two inputs in some fashion. 

Here, we present the first structure fit  in 
quasicrystals, in which energy and diffraction
information are combined in the same objective function.
Previously, powder diffraction data on $i$-TiZnNi \cite{hennig} was
systematically refined with energy inputs. 
In that work, 
each type of information was used separately
to fix the degrees of freedom which are most sensitive to it:
a least-squares refinement of diffraction data that
determined the species occupying each site type
was alternated with ab-initio calculations to relax the positions. 

In the present study we calculate energetics from pair potentials
microscopically derived using ``generalized pseudopotential 
theory''~\cite{GPT}, a total of six functions
(e.g. $V_{\rm AlNi}(r)$) for all combinations of the species.
The prominent feature of these potentials is Friedel oscillations, 
so that a typical  potential has three minima before the
cutoff radius (which we chose to be $10$~\AA).
The atom sites are called $\{ \R \}$, with 
$\chemsite(\R)$ designating the chemistry (Al, Ni, or Co)
of that site.
A convenient diagnostic tool in the refinement
of a decoration structure~\cite{AlMnI} is the ``site energy'', 
the portion of total energy ascribed to interactions of site $\R$:
   \begin{equation}
     \Esite (\R) \equiv \sum _{\R'} 
    V_{\chemsite(\R)\chemsite(\R')}(|\R-\R'|)  .
   \end{equation}
Then the  total energy is 
   \begin{equation}
    E_{\rm tot} \equiv \frac{1}{2} \sum _{\R\R'} \Esite (\R) .
   \label{eq:totalE}
   \end{equation}

Whereas previous refinements of quasicrystals
have used the hyperspace-cut representation of the structure, 
our results demonstrate the equal effectiveness of 
an alternative approach, the tile-decoration framework. 
The value of the tiling-decoration approach is that 
the structure models do not rely on periodicity or quasiperiodicity, 
and they are naturally represented in ``real'' space. 
Consequently, decoration is an elegant way~\cite{hennig} to transfer/combine 
information between the different system sizes appropriate for
different calculations, {\em e.g.}:
(i) numerical diffraction calculations from a large approximant;
(ii) eventual Monte Carlo simulations using a tile reshuffling update 
and a tile-tile Hamiltonian, in a smaller approximant;
(iii) molecular dynamics using pair potentials;
(iv) atom relaxation using an ab-initio total energy code, 
which is typically feasible only in  the smallest approximants.

The decoration formulation might be attractive for dealing
with {\it randomness} (an issue we ignore
in the present contribution, however).
A realistic structure  model is an 
{\em ensemble} of atomic configurations.  When the contents of 
different sites are independent random variables, then a
stochastic, mixed  occupation of each site is a fair
description.  If, however, the sites are highly correlated, 
a random tiling may be a more reasonable model; indeed, 
that is our present picture of the ``basic Ni'' phase of
$d$-AlNiCo~\cite{alnico}, 
A tile-decoration is superior
to a  hyperspace description in the random-tiling case:
all the randomness may be ascribed to the tile
configurations, while the decoration of each tiling is
deterministic. 
Some of the technical questions of a 
random-tiling  structure refinement were
discussed in Sec.5 of Ref.~\cite{rtdiff}.

\section{Structure modeling for combined fit}

To explain the decoration approach, we review nomenclature introduced
in \cite{AlMnI}: A {\it decoration} is a mapping which, given any
valid tile-configuration, produces a set of atom sites.  The sites are
grouped into {\it orbits}, each of which is {\it bound} to a
particular kind of tile, and lies in the same positions on every tile
of that kind.  An orbit plays a role like an orbit of symmetry-related
Wyckoff positions in a crystal structure.  In the hyperspace-cut
representation of a quasiperiodic structure, each member of the
decoration orbit on a representative tile would correspond to one
subdomain of the acceptance domain in perpendicular space.

More generally, orbits are bound to {\it tiling objects} which may be not
only tiles, but other geometrical constituents of the tiling such as
tile vertices -- e.g. we may choose to surround every vertex by an
identical symmetric cluster.  Another kind of a tiling object is
defined to overlay a certain local pattern in the tiling. We can then
introduce context dependence by {\it rebinding} certain atoms to the
new tiling object. Other atoms remain bound to the old objects, so we
do not define too many positional parameters.  One decoration rule is
a {\it rebinding} of another if it has more orbits, and the family of
structures produced from the original decoration (by varying the
parameter values) is a subset of the family produced from the
rebinding decoration.

By default we require that atomic decorations inherit the symmetry of
the tiling objects so that placing an atom at general position will
generate all symmetry equivalent sites. Furthermore, preservation of
the site symmetry constrains the degrees of freedom for atomic
displacements. In reality, forcing atomic motifs to obey the point
symmetry of tiling object is not always energetically favourable.

In the tile-decoration machinery,
such situations are handled by the concept of {\it symmetry-breaking}: 
a {\it flavour} (like an arrow on a tile)
is associated with each
(broken) symmetry related variant.~\cite{AlMnI}, so that the decoration
remains formally deterministic and reflects the symmetries
of the flavoured tiles.
The flavours may be initialized randomly, and in the simulation process
(for example Monte Carlo annealing) the symmetry-breaking may
be effectively used as an update move.

\begin{figure}
\epsfxsize=240pt \epsfbox{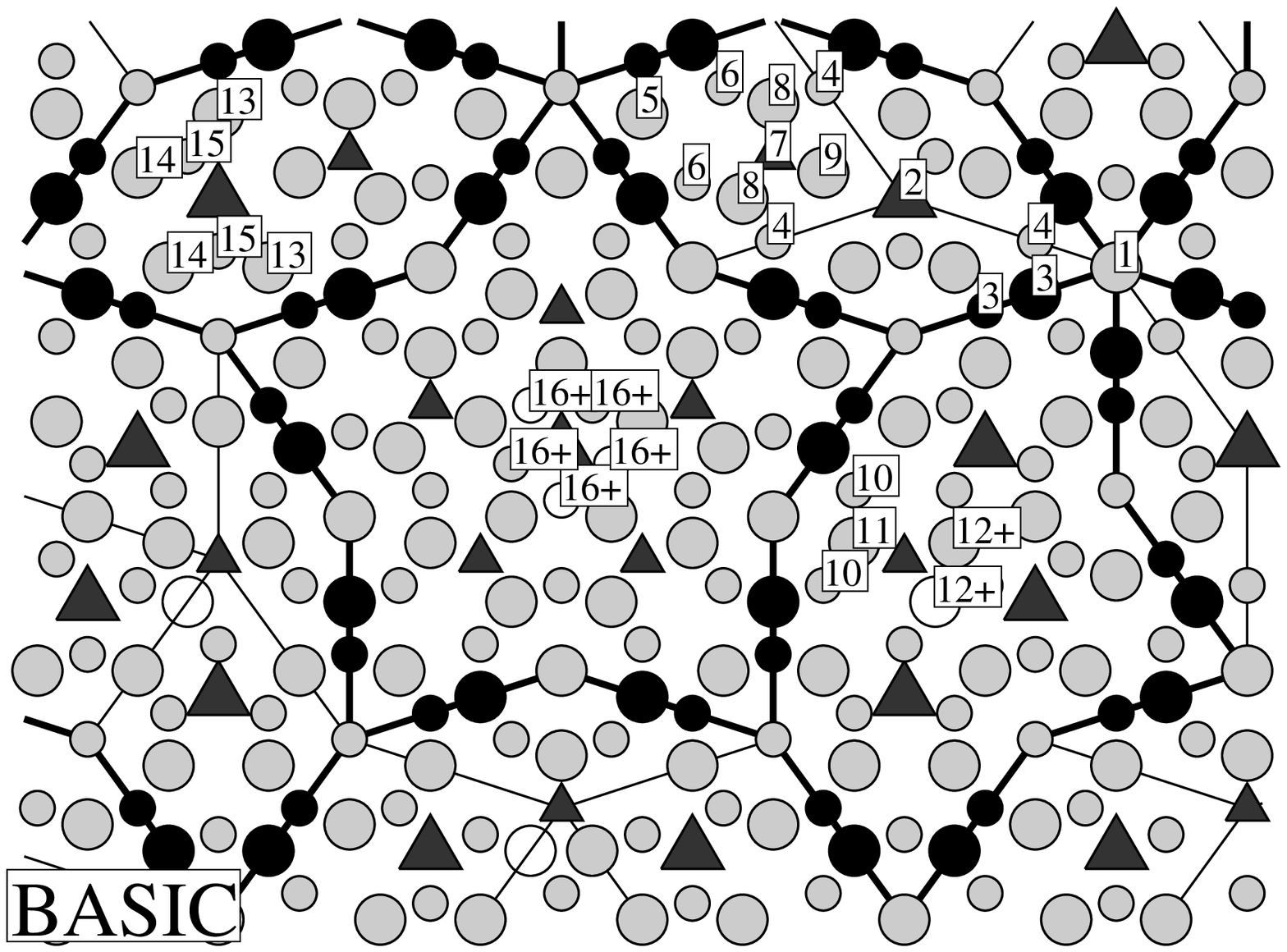} 
\epsfxsize=240pt \epsfbox{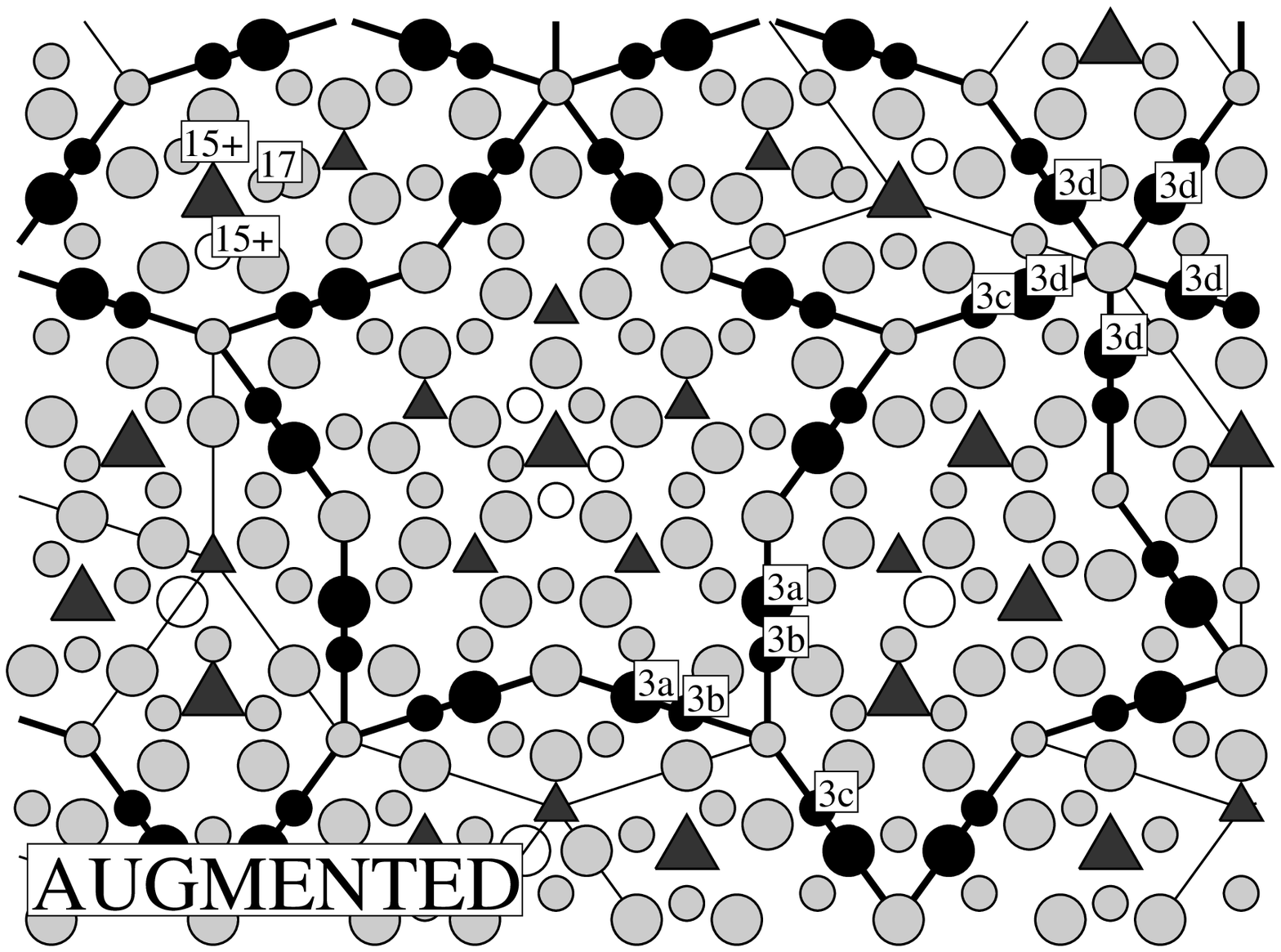} 
\caption{
Atomic decoration of the HBS tiling for ``basic'' (top panel) and
``augmented'' (bottom panel) bindings. Dark circles are Ni atoms,
dark triangles Co, gray Al. Unoccupied sites are empty circles.
Tiling orbits are shown as framed labels (once for each tiling
object). Thick lines are HBS-tiling edges, thin lines edges
of the inscribed rhombus tiling. For orbits that are binded
to HBS tiles, the edges of rhombuses inside the tiles are
not drawn.
Each symbol occurs in two
sizes, for upper and lower layer in the 4\AA\ thick slice
cut perpendicularly to the periodic direction.
}
\OMIT{MW: Why show some of the rhombi but not all?}
\label{fig:model}
\end{figure}

The decoration rule (see Fig.\ref{fig:model})
for our combined fit is adapted from the
deterministic model of~\cite{alnico}, in which atoms decorate an HBS
tiling with edge length $a_q\sim$6.4\AA
(similar to Fig.~3a of \cite{alnico}, but without the
complication of the ``bow-tie'' defects shown there.)
Atoms lie on flat layers that are spaced by 2\AA\ and are related
by the $10_5$ screw axis. Pairs of adjacent flat layers constitute
a ``4\AA\ bilayer'' which can be taken as the fundamental repeat unit
of a 4\AA\ periodic structure but can also be stacked in other fashions.
This model was
derived by successive elimination of the degrees of freedom in
Monte-Carlo simulations with the GPT pair potentials.

Notice that our Hexagon tile has a mirror plane running down its long
axis, but not transverse because the Al(9) atom breaks this symmetry.
We should visualize it as if marked by an
arrow to distinguish the two ends.  In the same way, the Boat's mirror
symmetry and the Star tile's fivefold rotation symmetry are broken
(respectively by Al(12) and Al(16) atoms) to
allow for our decoration.

Our ``basic'' rule uses as supplementary tiling objects the
Fat rhombi (which appear naturally when the HBS tiles are
subdivided into Penrose rhombi), as well as the interior 
vertices thereby created; this captures the similarity
of the local environments in all the 72$^\circ$ corners
of HBS tiles.
This rule is shown in Fig.\ref{fig:model}.
All Ni atoms are ascribed to the HBS
edges, Co atoms decorate the supplementary
interior vertices, as well
as a site on the long diagonal of every Fat rhombus.
Orbits 12 and 16 implement the broken symmetries of the Boat and Star.
In the end, this decoration rule associates 16 orbits with
nine tiling objects, and yields 38 positional parameters.

Refinement using the ``basic'' binding motivated a more complex 
``augmented'' binding (Table \ref{tab:orbits}), 
that proved to be more successful in fitting the data. 
Firstly, since the Ni(3) orbit alone contains more than 20\% of all
atoms, we split it among 3 objects, ``context-sensitive''
HBS Nodes. In both approximants we used, three classes of
HBS vertex stars occure: 5-fold, 3-fold and 2-fold (in random HBS,
also 4-fold and another 3-fold vertex type occurs). The symmetry
of these objects gives rise to 4 Ni orbits, labelled Ni(3$_a$-3$_d$).
The new orbits are listed in Table \ref{tab:orbits} below the
horizontal line.
Second, one of the two Al(15) atoms has been replaced by a vacancy, implementing
symmetry-breaking of the Hexagon mirror (Al(15+) in
Table \ref{tab:orbits}), and new atom Al(17) has been
placed near the center of the Hexagon.

The atomic density for the decoration rule (in either binding variant)
is 0.0705 atoms/\AA$^3$,
and the composition is Al$_{70}$Ni$_{20.7}$Co$_{9.3}$, slightly richer
in Co than the CHS sample (Al$_{70.6}$Ni$_{22.7}$Co$_{6.7}$). 
When the atoms are fixed at ``ideal'' sites, the minimal distance
occuring between Al-Al, Al-Co and Al-Ni pairs is 2.46\AA. The
structural energy of this initial model is $-0.297$ eV/atom for
a small approximant with 214 atoms per unit cell; using
positional degrees of freedom of the decoration rule it relaxes
to $-0.393$ eV/atom. Fully relaxed energy (atomic displacements
for atoms in the same orbit no longer equivalent) is $-0.422$ eV/atom,
and finally allowing atoms to relax their positions in a double
period (8\AA) unit cell, the energy drops to $-0.442$ eV/atom.

A drawback of the tile-decoration approach is that the infinite
non-periodic quasicrystal must usually be handled indirectly, being
approximated by an  periodic ``approximant''.
This necessarily has a 
``background phason strain'' which is, roughly, the
average tilt of the cut through hyperspace away from the irrational
orientation with 10-fold symmetry.)  
Consequently, each orbit of
reciprocal lattice vectors $\Q$ of the quasicrystal gets split into
several inequivalent orbits of the approximant's reciprocal lattice
vectors $\Q_{\rm app}$.  In the fit, we selected only one representative
$\Q_{\rm app}$ for each decagonal $\Q$, since the number of datapoints is
already large, but we monitored systematically the errors thus
introduced, calculating $\sigma_{\rm app}$ as the r.m.s. deviation 
of the difraction amplitudes among
all $\Q_{\rm app}$ mapping on a given $\Q$. 
For the larger approximant
discussed below, $\sigma_{\rm app}$ was similar to or less than the
experimentally observed uncertainty for all peaks.

A more systematic treatment of the approximant diffraction amplitudes
would be to symmetrize over all those deriving from a single orbit of
quasicrystal $\Q$. This would demand careful handling of the phase factors,
by properly centering the approximant structure both in real and in
perpendicular space~\cite{hennig}.

A further difficulty is that for smaller approximants, different
quasiperiodic $\Q$ vectors may map on the same approximant $\Q_{\rm app}$.
We found the smallest orthorhombic approximant avoiding such ambiguous
mappings for {\it all} reflections in the CHS data set has edge
lengths 61\AA\ $\times$ 32\AA\ in the decagonal plane.  In the
following, we label this approximant T$_{42}$ (for the number of HBS
vertices per unit cell).  For our decoration rule it contains 560
atoms per unit cell.  We found it satisfactory to use just one more
bigger approximant, with both sides in the decagonal plane larger by a
factor of $\tau\equiv(\sqrt{5}+1)/2$. We denote this tiling T$_{110}$
and decorate it with 1466 atoms.
\OMIT{MW: Marek, please check the number 1466 I inserted above}

\section{Procedure}

\OMIT{MW: I'm dropping the ``dataset'' and ``procedure'' labels because they
don't fit very well and don't seem necessary
{\it Dataset}. }
We used the CHS dataset~\cite{CHS} \OMIT{, crosschecked
with an older dataset from the same group~\cite{anc-oldref}}.
which contains 2767 unique reflections under the assumption
of 10/$m$ Laue group; by averaging over symmetry-related amplitudes,
to enforce 10/$m mm$ Laue symmetry,  
a set with 1544 unique reflections is obtained.
The two possibilities gave about equal internal $R$-factors.
Our structure model has $P10_5/mmc$ space
group, so we used the smaller 10/$mmm$ data set.
(Using the 10/$m$ set gave
practically the same results, except that typically both the R$_w$ and R factor
are increased by 1\%.)

The partial diffraction amplitude for tiling orbit $j$~is
\begin{equation}
\label{eq:Forb}
F^{\rm orb}_\orb(\Q)= \sum _\chem W_{\orb\chem}(\Q) f_\chem(\Q)
\sum_{\R\in \orb: \chemsite(\R)=\chem} e^{i\Q\cdot\R}.
\end{equation}
The inner summation in (\ref{eq:Forb})
runs over all atoms of species $\chem$ in orbit $\orb$, 
and $f_\chem(\Q)$ is the atomic form factor for that species.
The Debye-Waller (DW) coefficients
$\bxy_{\orb,\chem}$ and $\bz_{\orb,\chem}$ couple with
decagonal-plane and vertical $\Q$ components, and take
different values not only for different orbits, but
also for different species occupying the same orbit:
\begin{equation}
W_{\orb\chem}(\Q)=
    \exp[{-{1\over 4}(\bxy_{\orb\chem}\Q^2_{xy}+\bz_{\orb\chem}\Q^2_z}]
\label{eq:DW}
\end{equation}
The total calculated diffraction amplitude is then
  \begin{equation}
  \label{eq:Fcalc}
  F^{calc}(\Q)=e^{-{1 \over 4} b_\perp\Q^2_\perp}
       \sum_\orb{{|F^{\rm orb}|}_\orb(\Q)}.
\end{equation}
where we introduced a single perp-space DW coefficient $b_\perp$.
Finally, we compute goodness/reliability factors
\begin{equation}
\label{eq:chi}
\chi^2 \equiv \sum_{\Q}{(F^{obs}(\Q)-F^{calc}(\Q))^2/\sigma(\Q)^2}
\end{equation}
\begin{equation}
\label{eq:r}
R\equiv\sum_{\Q} \frac   {\big|F^{obs}(\Q)-F^{calc}(\Q)\big|}
                         {\sum_{\Q} F^{obs}(\Q)}.
\end{equation}
As usual, the weighted $R$-factor, called $R_w$, is
defined like (\ref{eq:r}) except that each term in the numerator
and denominator acquires a factor $1/\sigma(\Q)^2$.

{\it Implementation.}
We couple pair-potential energy with $\chi^2$ via parameter $\lambda$
in the objective function
for the non-linear least-squares minimization:
\begin{equation}
\label{eq:obj}
U = \chi^2 + (E-E^{\rm targ})^2/\lambda^2
\end{equation}
where $E$ is the energy (\ref{eq:totalE}), per atom.
Here $E^{\rm targ}$ is a ``target'' energy,~ \footnote
{When $\Delta E \equiv E^{\rm tot}-E^{\rm targ} \gg \lambda$, 
as is usually the case in the orbit-constrained relaxation, then 
$U= \chi^2 + 2 (\Delta E/\lambda^2) E^{\rm tot}$
would be practically equivalent to (\ref{eq:obj});
since the likelihood of parameters is supposed to 
behave as $\exp(-U)$,  $\lambda^2/2\Delta E$ is 
evidently playing  the role of a temperature.}
and is treated in the fit 
as a dummy datapoint with an ``error bar'' $\lambda$.
$E^{\rm targ}$ may be set to the total structural
energy per atom after an (unconstrained) relaxation $\Eunrel$.

Our fitting form 
(Eq.~(\ref{eq:DW}))
allows independent DW 
parameters $\bxy$ and $\bz$
for each orbit/species combination, 
but these are underdetermined by the data; to obtain sensible
results, the fit 
needs to be biased towards having similar DW factors
for similar atoms, in the spirit of the maximum-entropy method.
We implement this by adding terms for $\bxy$ and $\bz$
to the $\chi^2$ sum:
before each iteration, we calculate the average DW factor for
each species and the $xy$ and $z$ components of DW factor,
and set them as datapoints for the next iteration, with appropriately
chosen $\sigma_{DW}$. 

\section{Results}

Fitting the ``basic'' binding to the CHS dataset
we obtained reasonable, but not satisfactory R-factors $R_w$=0.122 
and $R$=0.202. The large displacement of atom Al(15) (see Fig.
\ref{fig:fmodel}, top) and other considerations motivated our
``augmented'' binding; Table  \ref{tab:summary} summarizes our
results for the latter.
\OMIT{(This rebinding has already been described above
under ``Structure modelling''.)}
The symmetry-constrained relaxed energy $\Erel$
of the 
``augmented''
rule is tolerably higher than $\Erel$
of the ``basic'' rule (see Table \ref{tab:summary}). 
However, the combined fit improved dramatically, displacing the
Al(15$_a$) atom towards the long body diagonal of H tile.
We have selected two structure solutions by the combined fit,
one with lowest R$_w$ factor but rather high energy, and one
with slightly worse R$_w$ factor, but comparable in energy
with the basic decoration rule (Table \ref{tab:summary}). 

The $F_{calc}(\Q)$ from our fit is compared to the
$F_{obs}(\Q)$  from the diffraction data in Fig. \ref{fig:fit}.

The fitted DW factors, especially those of Al, 
exhibit high degree of anisotropy (Table \ref{tab:summary}\ ).
This agrees with the 
TYT refinement~\cite{TYT} of a different sample of AlNiCo.
(The TYT sample was slightly richer 
in Al and Co than the sample for the CHS data~\cite{CHS} 
that we are fitting here.)

Most of the TYT hyperatom orbits correspond to our HBS tile
decoration orbits, 
and their individual orbit DW factors are mostly 
consistent with ours for the corresponding atoms.
The most pronounced disagreement is just the orbit 
Al(15+) for which $b_{z}/\bxy \approx 5$ in our refinement, 
but $\approx 1/5$ in the TYT refinement (see Table 2 in Ref.
\cite{TYT}, site label 9). This atom is at the same time
displaced farthest from its ideal position for both
TYT and our refinements, and the displacements have similar 
magnitude and direction for TYT and our ``{\em basic}'' binding
(see Fig.\ref{fig:fmodel}).
In the ``augmented'' binding, Al(15+) occurs in a completely
different environment, displacing towards long diagonal of H tile,
which explains the discrepancy in the DW factors.

\OMIT{MM-dec15: fixed/clarified above
This atom  was suffering largest displacement in our ``basic''
binding (see Fig.\ref{fig:fmodel}, top panel); and the displacement 
had similar magnitude and direction compared with TYT.
This discrepancy
can be easily understood, since our rebinding entirely
changed the environment of this atom (which also displaced significantly
parallel to the long axis of the hexagon tile).
[@x9xa ``WAS displaced on the mirror plane'' DID I MISUNDERSTAND?]
@MM: well - it displaced ON the long diagonal of the H tile (this is
clearly seen in the figure).
CLH: So you intended ``displaced ALONG''?
But no, orbit 15 does NOT lie on the long diagonal.
Do you mean it displaced  TOWARDS the long diagonal???]
MM: YES
}

\OMIT{MM-15dec: merged para below into the text above, and hopefully
clarified the sense.@@
Furthermore, in the TYT refinement it is precisely this atom which
is displaced the farthest from its ideal position;
this can be compared with the displacement
of Al(15) in our ``basic'' binding (see Fig.\ref{fig:fmodel}, top panel);
indeed, the displacements have similar magnitude and direction.
}
\OMIT{
[@x9c I didn't get the logic here: to explain why you get a DIFFERENT
anisotropy of DW than TYT, you argue that the atoms
have displaced in the SAME way. CAN YOU FIX TEXT AS YOU SAY?
@MM: I forgot to tell - the displacements were the same for the BASIC
binding, when there was pair of Al(15) per H tile; in the ``augmented''
binding, one of the two sites was left vacant(symm.-break) and an
EXTRA atom Al(17) has been placed near the center of H tile.
This Al(17) in just 2.3A distant from the Al(9); TYT model has the
same short pair of Als near the center of H tile. Another issue related
to this which I didnt discuss are partial occupancies in the TYT model.]
}

The TYT model and our decoration (in either variant) 
differ from the CHS model in having a larger
space group symmetry (10-fold screw axis vs. a 5-fold axis, 
4\AA\ translation vs.  8\AA\ translation), 
but also in other respects: higher atomic
density, and absence of transition metals (TM)
in the sites corresponding to hyperatom ``B''
in Ref.~\cite{CHS}.
Our  decorations are
more constrained than the other models, in that
mixed Al-TM occupancy is disallowed, partial 
occupation appears only through symmetry-breaking, 
and unphysical close pairs never appear on account
of the energy term in Eq.(\ref{eq:obj})
The CHS model achieved a lower weighted R-factor ($\sim$0.06) 
than the present refinement of the same data set,
at the price of including many more parameters. 
Enlarging the set of fitting parameters within our combined
energy-diffraction approach should further reduce the R-factor while
maintaining our energy-based tile decoration description.

\section{Discussion}

The rearrangement of the atomic decoration inside the Hexagon
tile --  driven, we think, by the diffraction term in (\ref{eq:obj})
-- is reminiscent of the atomic shifts observed in 
molecular dynamics (MD) simulations under the same 
potentials with an initial atomic configuration similar
to the present model~\cite{sendai}.
In those simulations, using supercells 8\AA\ or 16\AA\
thick in the stacking direction, slow
cooling led to a more favourable arrangement of atoms, in which 
one Al was pushed from the Al(15) position towards the center
of the Hexagon, while the other moved onto the long body diagonal
of the Hexagon. However, in contrast to our present model
in which two such atoms get squeezed into a column, in 
\cite{sendai}
only one atom per 8\AA\ period was displaced in this fashion, 
so that the column in the Hexagon center contained
three Al atoms per 8\AA\ period.
A recent MD study~\cite{gaehler}
confirms that the optimal Hexagon decoration has 8\AA\ period,
with atomic arrangement topologically equivalent to the
result of our combined fit, but
in \cite{gaehler} the atomic sites in the second 4\AA\ bilayer 
are {\em flipped} in each Hexagon 
across a mirror plane perpendicular to its body diagonal.
\OMIT{MW:12-20-02 Added sentence below to define ``bilayer''}
\OMIT{That is,  in \cite{gaehler} the
Hexagon has a glide plane perpendicular to the layers.}
\OMIT{Independent MD simulations, using our structure  
and potentials suggest that Al atoms can cyclically permute
in troughs around a Co atom or around the entire decagon.}

These findings again call for
study of the so-called ``stacking--fault'' mechanism,
in which the decoration rule would
remain essentially a 4\AA\ rule, and the variations
in the decoration from one 4\AA\ bilayer to the next would be 
described by ``flips'' of the tiles, occuring independently
in each bilayer.

\begin{figure}
\epsfxsize=240pt \epsfbox{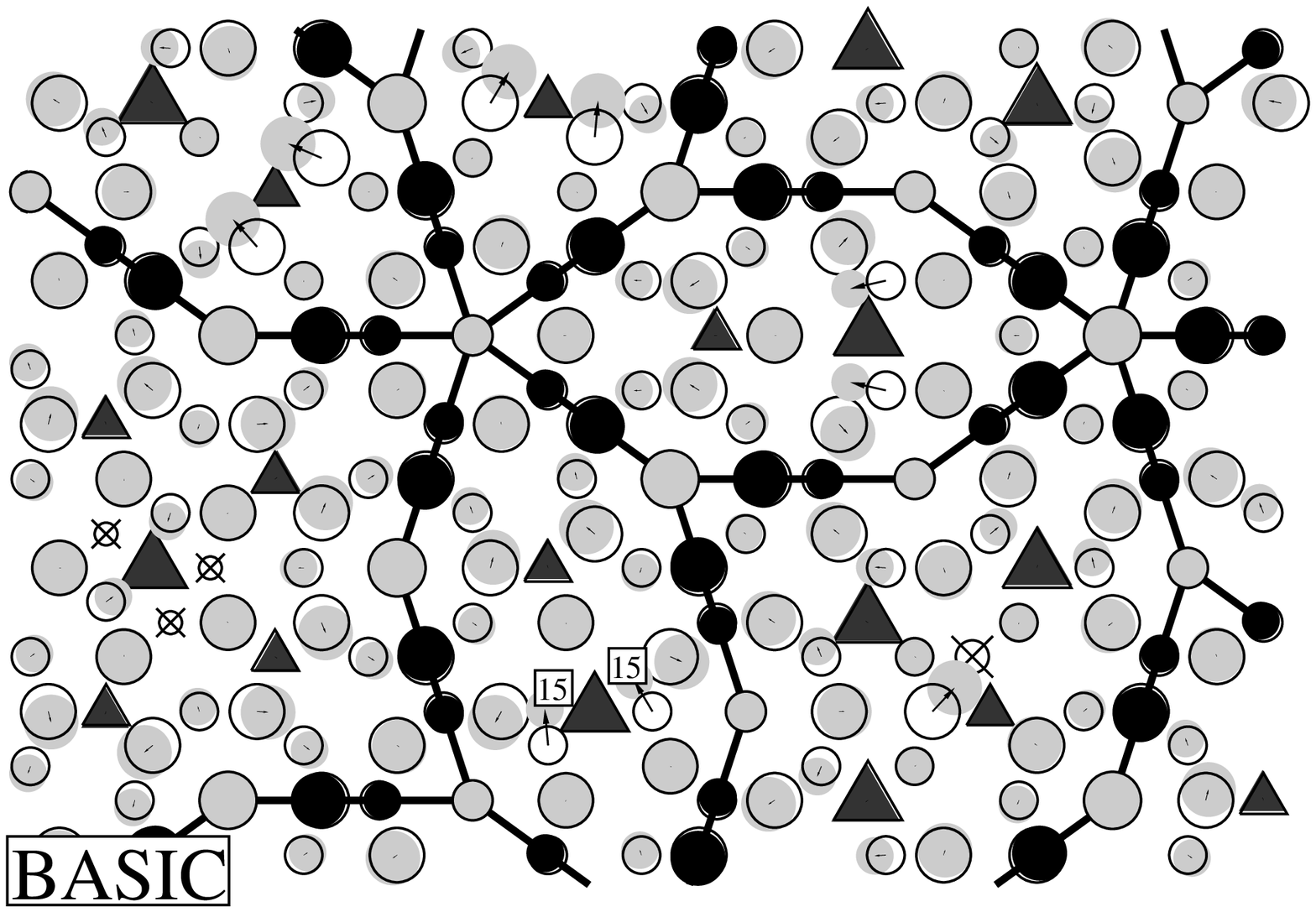} 
\epsfxsize=240pt \epsfbox{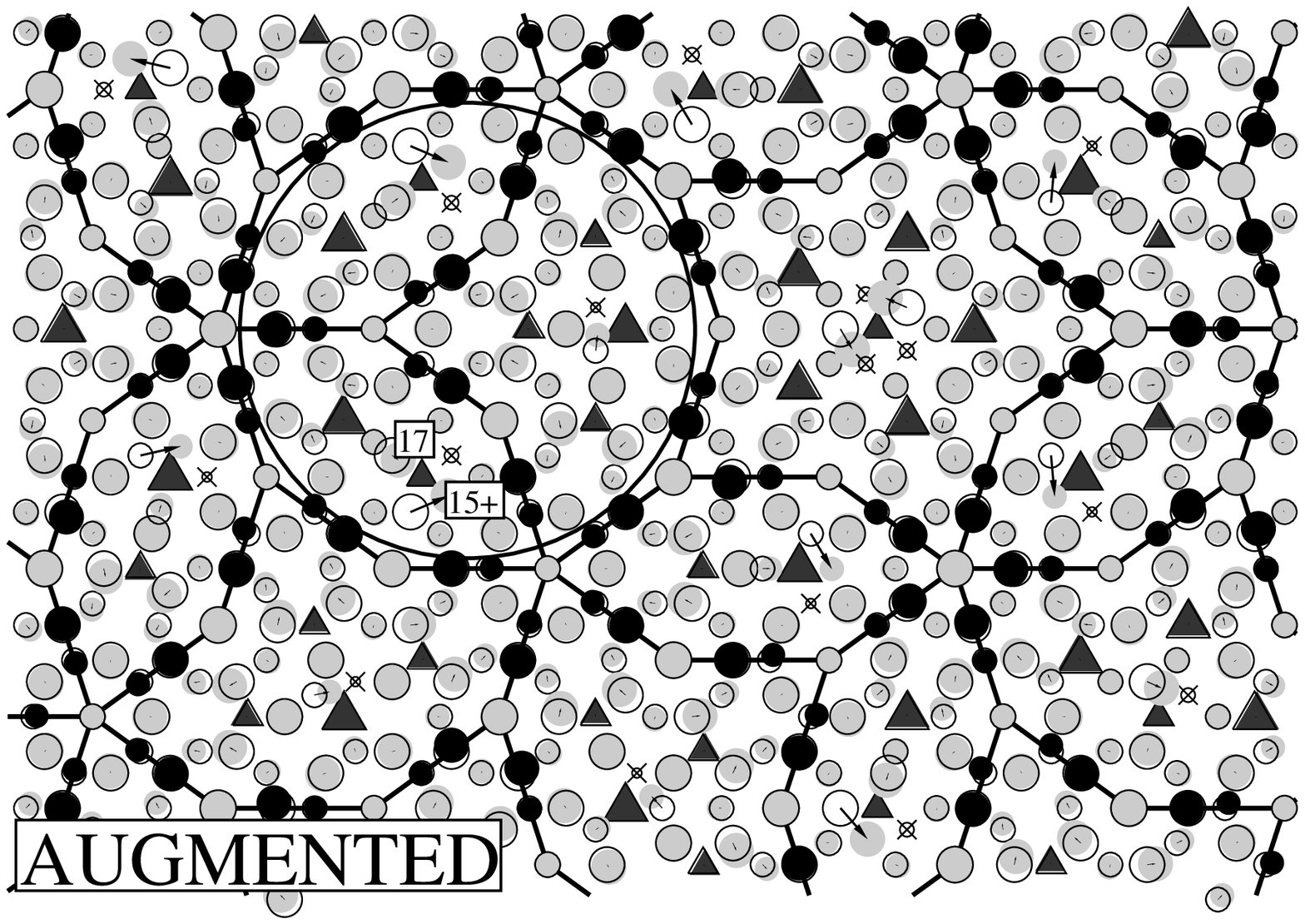} 
\caption{
{\it Top:} ``basic'' decoration rule, refined model.
Arrows indicate displacements (empty symbols stay at 
the ``ideal'' starting position) that occured during combined fit.
Vacant sites are shown as small circles and marked by crosses.
{\it Bottom:} 
``augmented'' binding, refined model. 
One decagonal columnar cluster is marked by two dashed circles,
near the outer shell of atoms, and near the inner ring of 10 TM
atoms.
}
\OMIT{MW: Why not label some Al(12) atoms in the top figure? You may
need to enlarge the bottom figure for clarity}
\label{fig:fmodel}
\end{figure}
\begin{figure}
\epsfxsize=240pt \epsfbox{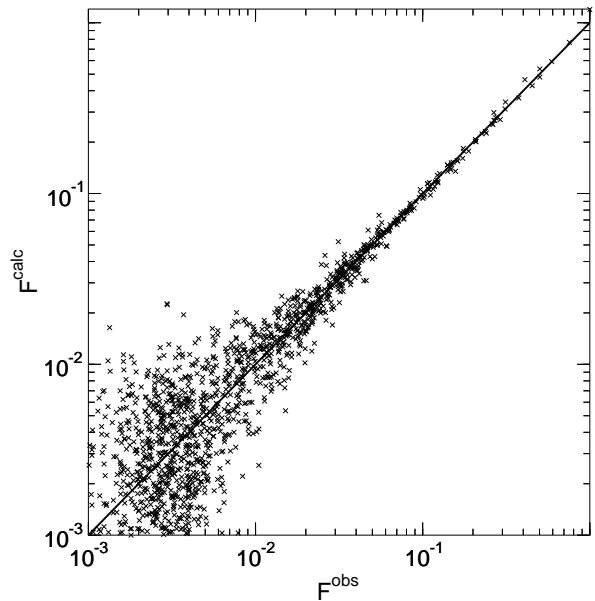} 
\caption{
Observed vs calculated diffraction amplitudes for approximant T$_{110}$,
final refined model of the ``augmented'' binding.
}
\label{fig:fit}
\end{figure}

The discussed variations in the atomic structure also
illuminate a motif  generally considered important in
decagonal structures: the 20\AA\ diameter decagon.
One such cluster in marked in Fig.~\ref{fig:fmodel} by the inscribed
circle.
Note the inner ring containing $\sim$20 Al atoms (Al(17) and Al(9))``squeezed''
in between 10 TM atoms (typically Co,  sometimes Ni), 
a feature that emerges in many HREM/HAADF
images~\cite{abe,yan}.
\OMIT{MW: I reworded this section
This could  appear as an optimal configuration in a 4\AA\  bilayer
{\em only} when energy and diffraction data are combined. } This is
energetically unfavorable in a structure with strict 4 \AA\
periodicity due to crowding of the repulsive Al atoms. It occurs in
our combined fit for a 4\AA\ bilayer {\em only} because it is favored
by the diffraction data.  On the other hand, as suggested by the MD
studies, it {\em should} emerge as a low energy configuration from
simulations performed for stacked bilayers which relax the strict 4
\AA\ periodicity.

In conclusion, we used tile-decoration machinery to set up a combined 
energy--diffraction data fit
of the d-AlNiCo structure. 
The resulting R-factors are
comparable to those from a previous study~\cite{CHS},
while using fewer parameters, and not allowing ad-hoc averaged/mixed
site occupancies. Apart from the inclusion of energies into
the diffraction fit, at this stage our refinement is completely
analogous to the hyperspace refinements, since we used strictly
approximants of the ideal, unperturbed quasiperiodic tiling.
The possible disorder correlations hidden in the experimental
data are still lumped into global parameters of the fit, 
mainly the perp-space
DW factor. 

A future prospect is to take full advantage
of the real-space formulation of the problem and recover the 
correlations in atomic occupancies directly from a ``grand
combined fit''. 
This would use Monte Carlo annealing to sample
all relevant degrees of freedom (swapping individual atoms,
flipping symmetry-broken tiling objects) in conjunction with
optimizing R-factors fitting the diffraction data.~\cite{rtdiff} 
If the pair potentials prove indeed sufficiently realistic,
the fit parameters may be further constrained by DW factors
pre-calculated from zero-T phonon spectra~\cite{phonon}, 
or extracted from room temperature MD annealing runs.

Finally, we expect our approach will prove optimal
for refinements of large quasicrystal approximant structures,
-- several such ``approximant''
phases are known to exist in the AlNiCo system~\cite{approximants} --
that are too complex to be handled by routine
crystallographic refinement approaches.
\OMIT{In such cases, the tile-decoration
approach to structure refinement has a clear 
technical advantage over the common hyperspace approach, 
since hyperatom division
into symmetrically related subdomains -- the essential ingredient of the 
hyperspace refinements -- and derivation of the displacive degrees of 
freedom for atoms would be tedious, if not arbitrary.} 

\vskip 0.1cm
{\bf Acknowledgments.}
We thank to W. Steurer for kindly  providing us his experimental
d-AlNiCo dataset, without which this work would not be possible.

\begin{footnotesize}
\begin{frenchspacing}

\end{frenchspacing}
\end{footnotesize}

\begin{table}
\begin{tabular}{llllll}
orbit & object &$x$ & $\bxy$ & $\bz$& $E_{site}$\\
\hline
Al(1) &  Node & 0.075 &  3.34 &  0.34 & 0.04\\ 
Co(2) &  Node$^r$ & 0.029 &  0.09 &  0.00 &$-2.21$\\ 
Al(4) &  Bond$^r$ & 0.104 &  0.92 &  2.79 &$-0.09$\\ 
Al(5) &   Fat & 0.064 &  0.00 &  4.73 &$-0.22$\\ 
Al(6) &   Fat & 0.128 &  2.15 &  2.45 & 0.24\\ 
Co(7) &   Fat$^r$ & 0.064 &  0.31 &  0.05 &$-2.28$\\ 
Al(8) &   Fat$^r$ & 0.128 &  2.91 &  3.25 & 0.49\\ 
Al(9) &   Fat$^r$ & 0.064 &  0.03 &  5.34 & 0.32\\ 
Al(10) &  Boat& 0.019 &  1.49 &  3.76 & 0.01\\ 
Al(11) &  Boat& 0.010 &  2.62 &  4.11 & 0.39\\ 
Al(12+) &  Boat& 0.010 &  0.00 &  0.21 & 0.65\\ 
Al(13) & Hex  & 0.030 &  1.23 &  3.63 &$-0.13$\\ 
Al(14) & Hex  & 0.030 &  0.00 &  3.82 & 0.06\\ 
Al(16+) & Star & 0.008 &  0.00 &  3.57 & 0.56\\  
\hline
Ni(3a)&Node$_2$& 0.079 & 1.05  &  0.40 &-1.34\\   
Ni(3b)&Node$_3$& 0.049 & 0.24  &  0.24 &-0.87\\   
Ni(3c)&Node$_3$& 0.025 & 0.92  &  0.36 &-1.55\\   
Ni(3d)&Node$_5$& 0.055 &  1.24 &  0.50 &-1.10\\ 
Al(15+) & Hex  & 0.015 &  0.82 &  5.37 & 0.61\\ 
Al(17)& Hex & 0.015 &  2.22 &  4.83 & 0.82\\  
\end{tabular}
\caption{
Atomic orbits in ``augmented'' binding. $x$ is fractional content, 
$\bxy$ and $\bz$ refined anisotropic DW coefficients, and $E_{site}$ 
are site energies in eV/atom, for
the approximant T$_{110}$. Orbits above the horizontal line were unchanged
from the ``basic'' binding.
Superscript $r$ denotes objects defined from the corners of the
vertices inside the HBS tiles (Fat$^r$ and Bond$^r$), see also
Fig. \ref{fig:model}
}
\OMIT{MW: We need a better definition of the column label ``x''}
\label{tab:orbits}
\end{table}

\begin{table}
\begin{tabular}{cccccc}
binding & $N_{\rm obj}$ & $N_{\rm orb}$ & $N_{\rm pos}$ & $N_{\rm par}$ & $\protect\Erel$ \\
basic & 9 & 16 & 38 & 72 & $-0.400$\\
augmented& 13& 23 & 58 & 106 & $-0.395$\\
\end{tabular}
\begin{tabular}{cccccc}
\hline
model & $b_{perp}$& $R_w$ & $R$ & $E$\\
T$_{42}$-basic  & 3.65 & 0.122 & 0.202& $-0.372$\\
T$_{42}$-augmented& 4.15 & 0.089 & 0.169& $-0.344$\\
T$_{110}$-augmented& 3.88 & 0.082 & 0.152& $-0.330$\\
 (smaller $\lambda$) & 3.93 & 0.086 & 0.159& $-0.370$\\
\end{tabular}
\caption{
{\it Top:} Numbers of tiling objects $N_{\rm obj}$, orbits $N_{\rm orb}$, 
positional parameters $N_{\rm pos}$ and total number of fit parameters
$N_{\rm par}$ (including anisotropic DW factors and the global $b_\perp$
DW factor). The last column is the energy 
per atom after constrained relaxation, without use of 
diffraction data.
{\it Bottom:} summary of the results for the combined fits, 
giving the perp-space Debye-Waller coefficient, 
the weighted ($R_w$) and unweighted $R$ factors.
The last row
reports an alternative refinement result for approximant T$_{110}$
with the ``augmented'' decoration rule, 
in which the $\lambda$ parameter in Eq. \ref{eq:obj}
was kept smaller to achieve lower energy.
}
\label{tab:summary}
\end{table}

\end{document}